# Engaging with Researchers and Raising Awareness of FAIR and Open Science through the FAIR+ Implementation Survey Tool (FAIRIST)

Kirkpatrick, Christine R.[1,2], Coakley, Kevin[1,3], Christopher, Julianne[1], Dutra, Inês[2]
[1] San Diego Supercomputer Center, University of California San Diego, La Jolla, CA USA
[2] Department of Computer Science, Faculty of Sciences of University of Porto, Porto, Portugal
[3] Department of Computer Science, Norwegian University of Science and Technology, Trondheim, Norway
christine@sdsc.edu, kcoakley@sdsc.edu, jchristopher@sdsc.edu, ines@dcc.fc.up.pt

**Abstract:** Six years after the seminal paper on FAIR was published, researchers still struggle to understand how to implement FAIR. For many researchers FAIR promises long-term benefits for near-term effort, requires skills not yet acquired, and is one more thing in a long list of unfunded mandates and onerous requirements on scientists. Even for those required to or who are convinced they must make time for FAIR research practices, the preference is for just-in-time advice properly sized to the scientific artifacts and process. Because of the generality of most FAIR implementation guidance, it is difficult for a researcher to adjust the advice to their situation. Technological advances, especially in the area of artificial intelligence (AI) and machine learning (ML), complicate FAIR adoption as researchers and data stewards ponder how to make software, workflows, and models FAIR and reproducible. The FAIR+ Implementation Survey Tool (FAIRIST) mitigates the problem by integrating research requirements with research proposals in a systematic way. FAIRIST factors in new scholarly outputs such as nanopublications and notebooks, and the various research artifacts related to AI research (data, models, workflows, and benchmarks). Researchers step through a self-serve survey process and receive a table ready for use in their DMP and/or work plan while gaining awareness of the FAIR Principles and Open Science concepts. FAIRIST is a model that uses part of the proposal process as a way to do outreach, raise awareness of FAIR dimensions and considerations, while providing just-in-time assistance for competitive proposals.

# 1. Introduction and Background

The FAIR Principles describe 15 aspirational dimensions of research data management (Wilkinson 2016). They provide a starting point for mapping out data stewardship practices needed for any given research project. However, there is no way to ensure all research objects adhere to FAIR, and FAIR is not all encompassing. For example, FAIR is silent on data quality and reproducibility. FAIR also does not comment on data sovereignty such as is covered in the CARE Principles of Indigenous Data Governance (Carroll 2022). Consensus is lacking on whose responsibility it is to ensure FAIRness of research objects. Stakeholders range from individual researchers to institutions to funders and publishers. In each case, the stakeholder group is looking for clear and practical advice and is less interested in philosophizing about the need for research data management practices or complex and detailed arguments over which approach is better. Researchers must address these topics only as much as funders require it. The US National Institutes of Health (NIH) requires a data management plan and, effective in 2023, increases the requirements to cover data sharing (National Institutes of Health). The National Science Foundation (NSF) requires a data management plan to be included with proposal materials. Funders around the globe require discussion of research data and FAIR in varying degrees. Countries and regions that lead the trends include the European Union's (EU) research funding calls, and Australia. Were it not for these funder requirements, researchers would only take these steps on a voluntary basis. However, this requirement provides a key moment during proposal preparation for outreach and awareness of the FAIR principles and how they relate to newer technologies. The FAIR+ Implementation Survey Tool (FAIRIST) creates information that can be included in a proposal's data management plan and/or the project description. Its contribution and value are as much in what it produces as well as the conversations and decisions it suggests by completing it. Even where support and services are not available to researchers from their institution, the mention of them can initiate important discussion.

This work is organized as follows. We introduce definitions and terminology in Section 2. Related work is also presented in this Section. In Section 3, we present the motivations and stakeholders of FAIRIST. Section 4 provides detail on FAIRIST's design and functionality. We close this work with a discussion and perspectives on future work.

## 2. Literature Review, Definitions, and Terminology

This work refers to concepts from information and computer science.

*FAIR* data is data that meets principles in four categories: Findability, Accessibility, Interoperability and Reusability. The **FAIR Principles** are the 15 principles that correspond to making research objects FAIR (Wilkinson 2016). The principles are not prescriptive and are not rules, but touchstones for concepts that lead to research object, or data, usability. According to libguides for FAIR data, the first thing to be in place to make data reusable is the ability to find them. Data should be easily findable both by humans and computers. Automatic and reliable discovery of datasets and services depends on machine-readable persistent identifiers and metadata. Persistent identifiers are important because they unambiguously identify your data and facilitate data citation. An example would be a Digital Object Identifier (DOI). The (meta)data should be retrievable by their identifier using a standardized and open communications protocol, with restrictions in place if necessary. Also, metadata should be available even when the data are no longer available. Data does not need to be all open, they can be restricted and still be FAIR. Open or not, data should be stored somewhere safe for the long-term. The data should be able to be integrated with other data, applications, and workflows. The format of the data should therefore be open and interpretable for various tools. The concept of interoperability applies both at the data and metadata level. Common formats and standards, and controlled vocabularies should be used. Ultimately, FAIR aims at optimizing the reuse of data. To do this, data should be well-documented, have a clear license to govern the terms of its reuse, and provenance information.

***FAIR Digital Objects:*** Even though the original FAIR Principles publication called out the need to make all types of research artifacts FAIR, there has been an overemphasis on data, e.g., a chunk of information or a single data point. This work acknowledges the need to make all digital objects FAIR, including software, models, algorithms, and workflows. The term 'FAIR Digital Object', or FDO, describes a concept and associated, evolving guidelines for packaging metadata about each chunk of information and the data together, and associating each component with its own unique identifier. The complete FDO is assigned a master identifier to the assembled package of data and metadata (De Smedt 2020). FAIRIST takes the approach that all research objects should be assigned identifiers and aims to move towards recommendations that provide advice to create FDO-compliant research objects.

***FAIR+***: To provide a shorthand for FAIR and reproducibility, FAIR+ is used as a term in this work.

***Open Science*** has been used by different stakeholders to focus on different aspects of openness from technological architecture to (public) accessibility of knowledge creation to measurement to democratization of access (Fecher 2014). This work focuses on the qualities of research processes that lead to openness including transparency and reproducibility through technical and practical approaches. FAIRIST supports open science aims via recommendations for implementing the FAIR Principles that relate to findability and accessibility, as well as reproducibility.

***Reproducibility***: This work uses Gundersen's definition of reproducibility, chiefly that science should be able to be reproduced not so much that the results are numerically identical, but so that the results support the same inferences drawn from the original research (Gundersen 2018). Although often mistaken as the "R" in FAIR, reproducibility is aided by implementation of the FAIR principles, especially those that pertain to openness of software, tools and libraries, accessibility of data, etc. Both FAIR and reproducibility are continuums more than a destination. Research reproducibility can be resource intensive, therefore researchers should do as much as possible to document and provide a path for another researcher to retest their conclusions, but it is understood that it is often not possible to recreate the exact same environment or to provide compute and storage needed for reproducibility work.

## Related Work

The FAIR principles and all literature give very good recommendations and guidance on how to address data in research projects. But usually, researchers work with text files and other non-structured documentation. Some efforts have been made to transform these recommendations and guidance into a computational tool that can standardize the process of data "FAIRification". Other important advancements that relate to FAIRIST include tools for interviewing researchers about FAIR implementation, data management practices, and upcoming tools for publishing and reusing data management plans.

***Argos:*** A joint effort between OpenAIRE and EUDAT, this platform provides a way to create and manage DMPs (Argos). Argos allows for manual entry or guidance with a wizard. Argos aims to document a research project and its outputs, mainly datasets. The Argos UI is streamlined and appears to employ modern UI/UX principles. It uses FAIR principles in how it collects data, utilizing APIs wherever possible so that researchers, institutions, and funders are not manually entered but connected to a unique identifier. The Dataset feature allows a researcher to document a dataset manually or prefilled from templates, customized for the needs of the funder. Argos allows for collaborative writing, and DMP templates can be added, updated, and modified. Argos provides DOI and DMP versioning via Zenodo and supports export in JSON format. Argos does require knowledge of data management concepts to complete the forms. Argos is a potential partner for integrating some of the questions from FAIRIST, although it would be a significant expansion in scope for the platform.

**DMPTool:** This tool provided by the California Digital Library (CDL), was created several years ago to assist researchers with the creation of their data management plans to accompany proposals (Praetzellis 2019). The survey is comprehensive and updated regularly. However, the primary text is entered by researchers into many text boxes. This can be daunting for researchers who are new to research data management and aren't sure where to start. As well, in practice researchers may only use the DMPTool once and then reuse plans from project to project, with minimal adjustment to the plan for the new project's needs. The survey and output of FAIRIST could be appended to the DMPTool and combined for ease of use by researchers.

**FAIR Implementation Profile (FIP) Wizard:** Created by members of the GO FAIR International Office and GO FAIR Foundation, the FIP wizard eases the creation of a FAIR Implementation Profile that can be read by machines (FIP Wizard, FIP Wizard Documentation). One answers questions in survey style, and the output is in the Resource Description Framework (RDF) format. The tool is also part of an exercise for communities to have discussions about (metadata) standards choices, such as has been used by the WorldFAIR project. Participants across several domains, or petals of the project, reported that having the discussion around metadata choices was as valuable as creating the FIP (Law 2022). The FIP Wizard employs some of the same techniques and design as for FAIRIST. However, it is concerned with aggregate information for a domain or subdomain of science, rather than an individual project's decision.

**FAIR Connect:** This new initiative from the GO FAIR Foundation and iOS Press, seeks to extend new tools for data stewards and researchers (FAIR Connect). It provides a way to publish FIPs and DMPs as nanopublications. It also allows for stewards to comment or endorse these inclusions. As well, it provides a way for stewards to be recognized for contributions via citations. FAIRIST outputs could be published in FAIR Connect as nanopublications and assigned persistent identifiers for citation and attribution.

## 3. Motivation and Stakeholders

Researchers desire practical advice for how to implement the FAIR principles, but are challenged by the steep learning curve and background needed to engage in data stewardship. Some may not even be aware of FAIR until they see it mentioned in a funding solicitation. A systematic tool can help by narrowing down topics based on research activities and outputs planned rather than the approach of presenting everything and leaving it to the researcher to select relevant principles. Such a tool should be designed to only broach topics that apply directly to the researchers' planned work. For those in the humanities who only plan to produce data and disseminate findings on a website, they won't encounter more complex topics such as where to share their machine learning models. Conversely, a computationally intensive project will find specific suggestions on where they might deposit ML artifacts and how to aid reproducibility of their work by others in the future.

FAIRIST began as a templated response used to assist colleagues in crafting FAIR and AI-aware data management plans (DMP). The implementation advice distilled as many of the FAIR principles as possible into a table that other researchers inserted into their DMP. Table 1 gives an example of the text created for an NSF proposal in astrophysics. This template was reused for other proposals where the project name was replaced and dimensions of FAIR were added or subtracted depending on the planned research. This capitalized on researchers' interest to learn more about FAIR implementation and research data management during the proposal process. However, proposal development is a very busy time and most attention is given to the project description or plan, not the DMP. In most cases, colleagues would use whatever guidance was provided. Knowing this, the advice given was created to be almost ready for inclusion in the DMP and areas to update were clearly marked in brackets (< >).

| FAIR Dimension | |
|---|---|
| **Findable** | <ul><li>Data will be assigned a PID <how?> and will be referenced on the Renaissance Simulation website https://rensimlab.github.io/</li><li>A catalog entry will be added to SDSC's FAIR Data Point.</li><li>Metadata and links to related ontologies will be available on the RSL website.</li><li>Where tags exist, schema.org descriptors will be utilized.</li></ul> |
| **Accessible** | <ul><li>Available via SDSC Cloud (object store), that doesn't require specialized software to access. This includes both the raw data and halo catalogs.</li><li>The surrogate and other ML benchmarks will be given to MLCommons for use by the Science Working Group and to be included in future benchmark distributions.</li><li>Any APIs will be versioned and described, linked from the RSL website.</li></ul> |
| **Interoperable** | <ul><li>Code stored on github (and linked from the RSL website)</li><li>Uses libraries from RSL that utilize xyz standard <or standard Python libraries, etc.>.</li><li>Uses standard references for astronomical bodies <more here>.</li><li>Uses <some standard way of describing shapes in geo specific way>.</li><li>Both input and output data are in HDF5 format.</li></ul> |

| Reusable | <ul><li>ML model and data will be deposited at OpenML.org</li><li>Notebooks will demonstrate how to assemble model and sample training datasets. Each notebook product will be assigned a DOI <using Zenodo or UCSD Library?></li><li>The RSL notebook interface is on github.</li><li>Provenance of the simulation creation will be available as part of the metadata.</li><li>We will post a notice designating all data as licensed under Creative Commons Attribution 4.0 International License.</li></ul> |
|---|---|

*Table 1*: Template that inspired the creation of FAIRIST

After filling out these templates manually a few times, it became clear that the process could be streamlined through a self-service survey. Even though every research project is different and the topics can be complex, much of the human logic could be distilled into if/then statements. For example, if the project will produce notebooks, then the DMP should specify where notebooks will be shared, if they will be given a DOI, and if a notebook template with metadata will be used.

FAIRIST provides customized text for a researcher to include in a data management plan or proposal. Some form of DMP is required by many federal funders. The added benefits of planning data management at the outset of a research project are many: it makes it easier to audit, to check compliance with requirements, and to document the project for helping both researchers and funding agencies. Raising topics as part of creating a required document can also put a research project in good stead for complying with other publication requirements later. For example, the Association for the Advancement of Artificial Intelligence (AAAI) hosts one of the most prestigious annual conferences for AI researchers. Papers submitted must also include a reproducibility checklist (AAAI Conference). Many of the implementation solutions to the FAIR principles aid researchers in also being ready for the reproducibility checklist. For example, for papers that rely on data sets, this question must be answered, "All novel datasets introduced in this paper will be made publicly available upon publication of the paper with a license that allows free usage for research purposes (yes/partial/no/NA)." Adherence to the FAIR principles that relate to clearly stating data usage licenses and accessibility of data would prepare researchers to answer "yes" to this question.

The stakeholders for a tool like FAIRIST include researchers from all domains and sectors, including academia and industry, although it is tuned for research grant proposals. Additional stakeholders include anyone involved in the proposal process where a DMP is required or discussion of the FAIR principles is beneficial. This could include cyberinfrastructure personnel, pre-award and project managers, and students or postdocs involved in proposal creation. This tool could also be used in synchronous and asynchronous trainings, such as the CODATA-RDA Schools of Research Data Science curriculum (CODATA-RDA-DataScienceSchools/Materials), a grantsmanship course (NSF HSI National STEM Resource Hub), or a data management plan training course hosted by a university library.

# 4. FAIRIST & Technology Used

FAIRIST surveys aspects of the project and then maps them to possible options and suggestions. This is a user-focused solution commonly used in applications such as tax preparation software or online dating or searching for a used car. The important qualities of this approach include: it makes a complex topic accessible, makes efficient use of researchers' time, and uses the time spent in the survey to lift awareness of the topic, its richness and dimensions. For example, if the project being described will produce Machine Learning (ML) models, then a follow up question is added asking, "Where will the ML models be shared?" with several options the user may or may not be aware of previously. An additional question asks,"What are the reproducibility considerations you will undertake to document analysis that utilizes ML?" (Figure 1). By providing check box options rather than only a free form text box, the user can gain knowledge about the topic that doesn't rely on specific understanding of FAIR+ concepts. The example shown in Figure 1 distills ML implementation factors that can affect reproducibility to introduce the concept and ways to remediate variability.

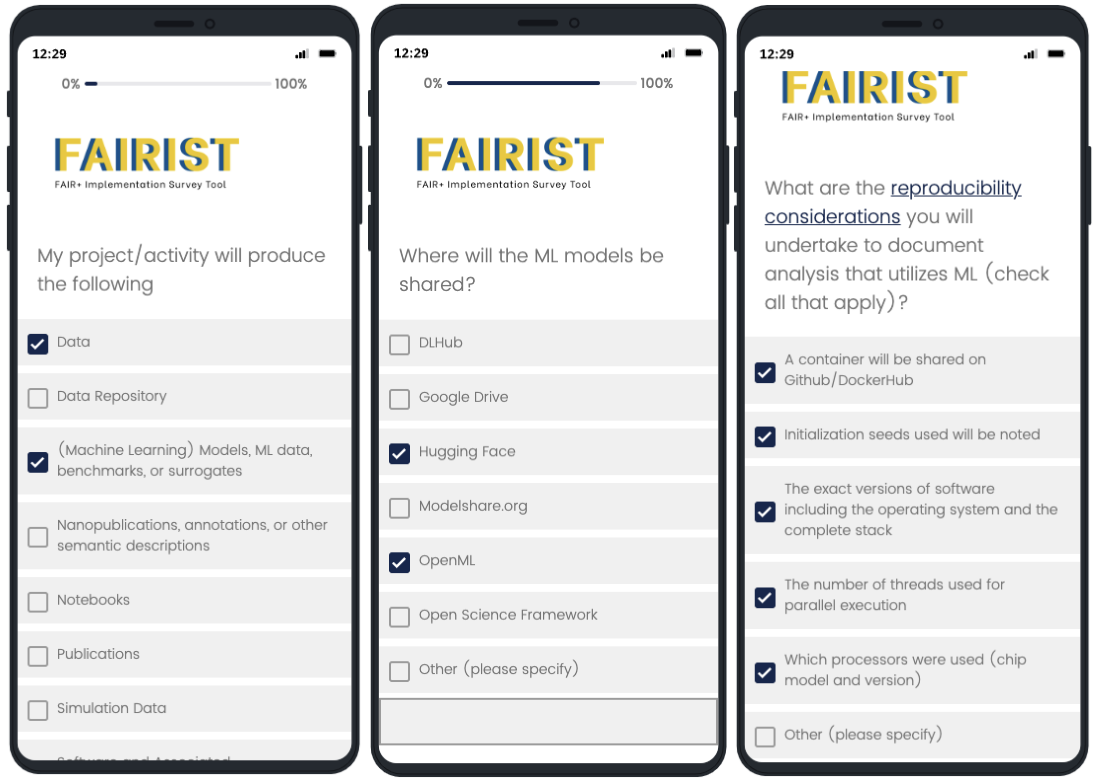

*Figure 1*: Embedded logic in FAIRIST expands the survey questions to fit the project described.

The reproducibility consideration options are distilled from a much longer and complex Computer Science paper on sources of irreproducibility (Gunderson 2022). The source paper is linked in the FAIRIST survey question, in case the user wishes to read more about the topic before deciding or implementing the suggestions. The advice from the paper was also adapted for a postcard-sized outreach material (Figure 2) that could be used for outreach and awareness building of the concept and FAIRIST tool. This approach could be used to rapidly put other research data management scholarship into practice.

*Figure 2*: Outreach and awareness building material that could be used in concert with FAIRIST by libraries, research computing facilitators, and other researcher support personnel.

FAIRIST was built in Qualtrics, a business survey tool with a full-featured user interface (UI) that allows for survey customization as the input is given (Qualtrics). For example, it is possible to ask up front what types of research objects will be created and add or skip questions automatically based on the first response. The Qualtrics UI allows for a non-programmer to refine the form iteratively, enabling an Agile approach. Qualtrics is limited in customized output. Some of the responses are used to determine what additional questions to ask, whereas the other responses set variables. Using the Qualtrics API, the variables call a Python script hosted on a local cloud instance that transforms the variables into completed sentences (Coakley 2022). This text output is formatted for inclusion in a DMP, but could also be extended to include a machine actionable DMP. This would be accomplished by providing the text formatted for written language as well as in RDF.

The Qualtrics API is limited and slow; it can take from 1-5 minutes for the output to be ready. A user is notified by email when the FAIRIST output is ready to be retrieved. The link URL uses a random string, so that other DMPs can't be viewed as would be possible if the key value was displayed in the URL. After a few minutes, an email is sent to notify the user that the recommendations are available for viewing. An example of the output is shown in Table 2.

# FAIRIST Recommendations

Based on your responses, the following recommendations are included for your consideration and/or inclusion in your project's Data Management Plan.

**Types of Data**

Research objects associated with the project can be classified into the following groups:

- Data
- (Machine Learning) Models

**Data Stewardship Practices Planned**

Table 1 shows specific data stewardship actions that will be undertaken during the project as they relate to the high-level goals of FAIR.

| FAIR Dimension | Research Data Stewardship Practices Planned |
|---|---|
| Findable | <ul><li>Research products will be posted to the Project website.</li><li>Data will be assigned a unique identifier per community best practices and will be referenced on the project's website.</li><li>Metadata and links to related ontologies will be available on the Project website.</li><li>Where tags exist, schema.org descriptors will be utilized.</li></ul> |
| Accessible | <ul><li>Available via open, web accessible folder.</li><li>All data is open.</li></ul> |
| Interoperable | <ul><li>Code stored on github (and linked from the Project website).</li><li>Uses libraries included with the code.</li><li>Both input and output data are in HDF5 format.</li></ul> |
| Reusable | <ul><li>ML model and data will be deposited at OpenML.org.</li><li>A notice posted will designate research objects as licensed under CC-BY.</li></ul> |

Table 1: Data Stewardship Practices Planned by FAIR Dimension

*Table 2*: Example Output from FAIRIST Showing Recommendations

While this is not a long-term solution, using Qualtrics allows for incremental improvements based on user testing and input from other experts. Once the FAIRIST's questions and output have been well tested and refined, FAIRIST should migrate to a custom, efficient, stand-alone Python script and/or be integrated with an existing DMP tool. Output would instead be available immediately on the confirmation screen.

A lesson learned for future tool developers, and borrowed from the Agile software development community, is to not get so bogged down in creating a web tool or new software, that the point in creating it is lost or delayed. It is reasonable to launch such a tool with the information known and to amend it as new knowledge is gained, treating it more as a living resource than a fixed resource published once.

FAIRIST is available for use at http://fairist.sdsc.edu/

# Conclusion & Future Work

While FAIR implementation is dependent on the specific factors for each research project and domain, as well as changes with evolving technology, it is possible to provide researchers with concrete advice. Tools such as FAIRIST, provide a framework for embedding new information as practices develop and raising awareness on open science practices. By utilizing Agile methodologies and readily available cloud-based tools to create FAIR tools and resources, implementation advice can be more rapidly distilled and presented to researchers. These takeaways can be packaged not only for use in FAIRIST and tools like it, but also reformatted as outreach and awareness tools that promote both tools and FAIR+ concepts. Though created with the motivation to assist researchers and their teams with FAIR implementation and to increase adoption of the FAIR Principles, survey tools with proactive suggestions can assist researchers in other ways. Streamlined tools like FAIRIST can anticipate publishing and other funder requirements.

Future work should include a wider review of the survey options and outputs by experts in information science and research computing. It would be more sustainable if these questions and survey techniques were adopted by an existing tool. However, if that does not occur, an advisory board should be formed to guide decisions. For example, one of the questions, "Where will your ML datasets be shared?" provides several options. Should the survey reflect the current practices or eliminate options the community determines to be suboptimal? If FAIRIST remains a stand-alone tool it should be converted to an application with a database for storing surveys and accessing past FAIRIST output.

There are numerous other sources to mine for potential questions and implementation suggestions. This work focused on computer science and AI because of the authors' backgrounds and a perceived gap in implementation advice. A future version of FAIRIST could include custom options tailored to advice for specific domains. This is beyond the complexity that can be handled by the current survey tool but would be possible in a future iteration of FAIRIST. For example, for projects that will produce a new domain repository and are from the Earth Sciences, FAIRIST could offer the option to include the repository in the Magnetics Information Consortium (MagIC) or Council of Data Facilities (CDF) consortium (Council of Data Facilities). At the moment, this option is offered to anyone, regardless of scientific domain, that indicates the project will create a domain repository. FAIRIST could be further tailored to research needs by asking custom questions based on the agency funding source. As funders or institutions implement machine actionable DMPs, FAIRIST could also include the implementation guidance in machine readable format, e.g., triples in RDF. This could then be used to automatically verify compliance with the planned research data management practices.

**Acknowledgements**
Thanks to Melissa Cragin (Rice University) for her early feedback on the FAIRIST survey questions and to Odd Erik Gundersen (NTNU) for his mentorship regarding AI reproducibility. The FAIRIST logo was created by Alexandra Andrieu (SDSC, UC San Diego). Lynne Schreiber (SDSC, UC San Diego) co-created the outreach materials shown in Figure 2.

**Funding**
This work was partially funded by National Science Foundation (NSF) awards #2226453, #1916481 and FAIR pilot funding support from the San Diego Supercomputer Center.

**Author Information**

**Christine R. Kirkpatrick** is the division director of Research Data Services at the San Diego Supercomputer Center, UC San Diego. She heads GO FAIR US and is Secretary General of CODATA.

**Kevin Coakley** is a Senior Cloud Integration Specialist at the San Diego Supercomputer Center, UC San Diego. Kevin is a PhD student in the computer science department at NTNU, specializing in AI reproducibility.

**Julie Christopher** is an IT project manager at the San Diego Supercomputer Center, UC San Diego.

**Inês Dutra** is an Assistant Professor in the Department of Computer Science, Faculty of Sciences of University of Porto, Portugal. Her main research interests are parallelism, logic programming and interpretable machine learning models.